\documentclass[aps, pra, preprint, superscriptaddress, amsmath, amssymb, showpacs]{revtex4-1}
\usepackage[english]{babel}
\usepackage[utf8]{inputenc}
\usepackage{longtable}
\usepackage{amssymb,bm}
\usepackage{graphicx}
\usepackage{amsmath}
\usepackage{epstopdf}
\usepackage{float}

\begin{document}
	
\title{Consistent analysis of $f_1 (1285)$ meson form factors}

\author{A.I.\,Milstein}
\email{milstein@inp.nsk.su}
\affiliation{Budker Institute of Nuclear Physics of SB RAS, 630090 Novosibirsk, Russia}
\affiliation{Novosibirsk State University, 630090 Novosibirsk, Russia}

\author{A.S.\,Rudenko}
\email{a.s.rudenko@inp.nsk.su}
\affiliation{Budker Institute of Nuclear Physics of SB RAS, 630090 Novosibirsk, Russia}
\affiliation{Novosibirsk State University, 630090 Novosibirsk, Russia}

%\date{\today}

\begin{abstract}
Parameterization of the form factors of $f_1 (1285)$ meson is proposed. This parameterization is consistent with the available experimental data on the cross sections of $f_1 (1285)$ meson production in the processes $e^+e^- \to f_1 (1285)$ and $e^+e^- \to e^+e^-f_1 (1285)$, as well as on the widths of the decays $f_1 (1285)\to e^+e^-$, $f_1 (1285)\to \rho^0\gamma$, $f_1 (1285)\to \rho^0\pi^+\pi^-$, and $f_1 (1285)\to 2\pi^+2\pi^-$. Our parameterization is also consistent with the predictions for the asymptotic behavior of these form factors.
\end{abstract}

%\pacs{12.20.Ds, 31.30.J-, 42.50.Xa}
\maketitle

%---------------------------------------------------------

\section{Introduction}
The experimental and theoretical investigations of two-photon production of $f_1 (1285)$ meson is very interesting since a particle with the spin $S=1$ cannot be produced in a collision of two real photons due to their identity \cite{Landau:48}. However, $f_1 (1285)$ meson can be produced in a collision of two virtual photons or one virtual photon and one real photon. Therefore, the probability of these processes can be sensitive to the $f_1 (1285)$ meson internal structure, i.e., to the dependence of the form factors on photon virtualities. At present, there are a few experimental \cite{Gidal:87, Aihara:88, Amelin:95, Achard:02, Dickson:16, Achasov:19} and theoretical \cite{Kopp:74, Kaplan:78, Kuhn:79, Renard:84, Cahn:87, Schuler:98, Kochelev:09, Wang:17, Wang:2017, Rudenko:17} results on production and decays of $f_1 (1285)$ meson. Unfortunately, QCD cannot predict now the shapes of the corresponding form factors at moderate photon virtualities. Some predictions for the form factors exist only in the region of very large virtualities, though even in this case the particular shape of the form factors depends on the unknown wave functions of $f_1 (1285)$ meson \cite{Kopp:74}. Therefore, to understand the features of $f_1 (1285)$ meson production processes, it is necessary to use the phenomenological parameterization of the form factors, which should be consistent with all available experimental data. This is the goal of our work.

%---------------------------------------------------------

\section{General structure of $f_1 (1285) \to \gamma^*\gamma^*$ amplitude}
Since $f_1 (1285)$ meson has positive parity, the amplitude ${\cal M}$ of the $f_1 (1285)$ meson decay into two vector particles with negative parities, momenta $k_1$ and $k_2$, and polarization vectors 
$e_1^\mu$ and $e_2^\mu$ has the form \cite{Kopp:74, Rudenko:17}
\begin{equation}\label{lag}
\begin{split}
{\cal M}&=\dfrac{i}{M^2}\epsilon_{\mu\nu\rho\sigma}e_1^{*\mu} e_2^{*\nu}\Big[F(k_1^2,k_2^2)k_2^\rho k_1^\sigma A^{\delta}(k_1-k_2)_\delta \\
&-k_2^2G(k_1^2,k_2^2)A^{\rho} k_1^\sigma+k_1^2G(k_2^2,k_1^2)A^{\rho} k_2^\sigma\Big]\,,
\end{split}
\end{equation}
where $M$ is the mass of $f_1 (1285)$ meson and $A^\rho$ is its polarization pseudovector, $F(k_1^2,k_2^2)$ and $G(k_1^2,k_2^2)$ are two dimensionless form factors, $e_\mu k^\mu=0$, $\epsilon_{0123}=-1$. Since the amplitude $\cal M$ should be symmetric with respect to permutation $1\leftrightarrow 2$, then $F(k_1^2,k_2^2)=-F(k_2^2,k_1^2)$. As should be, ${\cal M}=0$ at $k_1^2=k_2^2=0$.

Consider the $f_1 (1285)$ meson decay into two virtual $\rho^0$ mesons. In this case we represent the corresponding form factors $F_{\rho\rho}(k_1^2,k_2^2)$ and $G_{\rho\rho}(k_1^2,k_2^2)$ in the form
\begin{equation}\label{rhorho}
\begin{gathered}
F_{\rho\rho}(k_1^2,k_2^2)=\dfrac{\tilde g_1 M^3 (k_2^2-k_1^2)}{q}\,,\quad G_{\rho\rho}(k_1^2,k_2^2)=\dfrac{\tilde g_2 M^5}{q}\,, \\
q= \dfrac{1}{M}\sqrt{\nu^2-k_1^2k_2^2}\,,\quad \nu=k_1k_2=\dfrac{1}{2}(M^2-k_1^2-k_2^2)\,,
\end{gathered}
\end{equation}
where $\tilde g_1$ and $\tilde g_2$ are some constants. In the rest frame of $f_1 (1285)$ meson, one has $q=|\bm k_1|=|\bm k_2|$. The parameterizations in Eq.~\eqref{rhorho} are the simplest forms of the form factors $F_{\rho\rho}$ and $G_{\rho\rho}$, which provide the antisymmetry of the form factor $F_{\rho\rho}$ with respect to replacement $k_1^2\leftrightarrow k_2^2$ and the correct asymptotic behavior of the matrix element \eqref{lag} at large virtualities \cite{Kopp:74}.

For the $f_1 (1285)$ meson decay into virtual $\rho^0$ meson with the momentum $k_1$ and a virtual photon with the momentum $k_2$, we follow the logic of the vector dominance model and write the corresponding form factors as
\begin{equation}\label{rhogamma}
F_{\rho\gamma}(k_1^2,k_2^2)=\dfrac{(ef_\rho)\tilde g_1 M^3 (k_2^2-k_1^2)}{q(k_2^2-\mu_\rho^2)}\,,\quad G_{\rho\gamma}(k_1^2,k_2^2)=\dfrac{(ef_\rho)\tilde g_2 M^5}{q(k_2^2-\mu_\rho^2)}\,.
\end{equation}
Here $\mu_\rho^2=m_\rho^2-im_\rho\Gamma_\rho$, where $m_\rho$ and $\Gamma_\rho$ are the mass and the width of $\rho^0$ meson, respectively, $ef_\rho$ is the constant of 
the $\rho^0$ meson-photon transition. This quantity can be expressed via the width $\Gamma_{\rho \to ee}$ of the $\rho^0$ meson decay into $e^+e^-$ pair,
$$ef_\rho = \left(\dfrac{3\Gamma_{\rho \to ee}m_\rho^3}{4\pi\alpha}\right)^{1/2}\,,$$
where $e$ is the electron charge, $\alpha=e^2 \approx 1/137$ is the fine structure constant, and $\hbar=c=1$.

At last, the form factors for the amplitude of the $f_1 (1285)$ meson decay into two virtual photons read
\begin{equation}\label{gammagamma}
\begin{gathered}
F(k_1^2,k_2^2)=\dfrac{ g_1 M^3 (k_2^2-k_1^2)}{q(k_1^2-\mu_\rho^2)(k_2^2-\mu_\rho^2)}\,,\quad G(k_1^2,k_2^2)=\dfrac{g_2 M^5}{q(k_1^2-\mu_\rho^2)(k_2^2-\mu_\rho^2)}\,, \\
g_1=(ef_\rho)^2\tilde g_1\,,\quad g_2=(ef_\rho)^2\tilde g_2\,.
\end{gathered}
\end{equation}
In this formula, the quantity $q$ in the denominators has a kinematic origin and provides a correct behavior of the cross sections at very large virtualities as predicted in \cite{Kopp:74}.

Using Eq.~\eqref{lag} it is easy to derive the spiral amplitudes of the transition 
$\gamma^*\gamma^*\to f_1 (1285)$ in rest frame of $f_1 (1285)$ meson. In this frame
\begin{equation}
\begin{gathered}
k_1=(\omega_1,\bm q)\,,\quad k_2=(\omega_2,-\bm q)\,, \\
e_{1+}=(0,\bm e_{1+})=e_{2-}\,,\quad e_{1-}=(0,\bm e_{1-})=e_{2+}\,, \\
e_{10}=\dfrac{1}{\sqrt{-k_1^2}}(q,\omega_1\bm e_z)\,,\quad
e_{20}=\dfrac{1}{\sqrt{-k_2^2}}(-q,\omega_2\bm e_z)\,, \\
\bm e_{1+}=\dfrac{1}{\sqrt{2}}(-i\bm e_x+\bm e_y)\,,\quad \bm e_{1-}=\dfrac{1}{\sqrt{2}}(i\bm e_x+\bm e_y)\,,\quad \bm e_z=\bm q/q\,, \\
\omega_1=\dfrac{k_1^2+\nu}{M}\,,\quad \omega_2=\dfrac{k_2^2+\nu}{M}\,.
\end{gathered}
\end{equation}
Then we have for the spiral amplitudes
\begin{equation}\label{ampl}
\begin{gathered}
{\cal M}_{++}=\dfrac{ (\bm e_z\cdot\bm A^*)}{M^3}\Big[2q^2M^2F(k_1^2,k_2^2)-k_2^2(k_1^2+\nu)G(k_1^2,k_2^2)+k_1^2(k_2^2+\nu)G(k_2^2,k_1^2)\Big]\,, \\
{\cal M}_{--}=-{\cal M}_{++}\,,\quad{\cal M}_{+-}={\cal M}_{-+}={\cal M}_{00}=0\,, \\
{\cal M}_{+0}=\dfrac{ (\bm e_{1+}\cdot\bm A^*)\sqrt{-k_2^2}}{M^2}\Big[k_1^2G(k_2^2,k_1^2)-\nu G(k_1^2,k_2^2)\Big]\,, \\
{\cal M}_{-0}=-\dfrac{ (\bm e_{1-}\cdot\bm A^*)\sqrt{-k_2^2}}{M^2}\Big[k_1^2G(k_2^2,k_1^2)-\nu G(k_1^2,k_2^2)\Big]\,, \\
{\cal M}_{0+}=-\dfrac{(\bm e_{2+}\cdot\bm A^*)\sqrt{-k_1^2}}{M^2}\Big[k_2^2 G(k_1^2,k_2^2)-\nu G(k_2^2,k_1^2)\Big]\,, \\
{\cal M}_{0-}=\dfrac{ (\bm e_{2-}\cdot\bm A^*)\sqrt{-k_1^2}}{M^2}\Big[k_2^2 G(k_1^2,k_2^2)-\nu G(k_2^2,k_1^2)\Big]\,.
\end{gathered}
\end{equation}
Here subscripts ``$+,\,-,\,0$'' stand for the corresponding helicities of the virtual photons.
Using this formula and the parameterization of the form factors \eqref{gammagamma}, we can describe various processes, extract the parameters of the model, and compare our predictions with the experimental data available. 

%---------------------------------------------------------

\section{Parameters of the model}\label{P}
It is seen from Eq.~\eqref{ampl} that the amplitudes ${\cal M}_{\pm0}$ and ${\cal M}_{0\pm}$ are independent of the form factor $F(k_1^2,k_2^2)$.
Therefore, the experimental data, which are related solely to these amplitudes, allow us to extract the constant $g_2$. 

L3 Collaboration has measured the width $\Gamma^{TS}_{\gamma\gamma^*}$ of the $f_1 (1285)$ meson decay into one real photon and one virtual photon with the longitudinal polarization \cite{Achard:02}: 
\begin{equation}\label{gt}
\widetilde\Gamma_{\gamma\gamma}=
\lim_{k_2\to 0}\dfrac{M^2}{(-k^2_2)}\Gamma^{TS}_{\gamma\gamma^*}=3.5\pm 0.6\,(\mbox{stat.})\pm0.5\,(\mbox{sys.})\,\mbox{keV}.
\end{equation}
Using this result and Eq.~\eqref{ampl} we find
\begin{equation}\label{g21}
|g_2|=\sqrt{\dfrac{3\widetilde\Gamma_{\gamma\gamma}}{\pi M}}\left(\dfrac{m_\rho}{M}\right)^4=(2.2\pm 0.2)\cdot 10^{-4}\,.
\end{equation}
The uncertainty in this formula comes from experimental uncertainties of all quantities in it, though the main contribution is given by the uncertainty of $\widetilde\Gamma_{\gamma\gamma}$.

VES Collaboration has studied the process $f_1 (1285)\to \rho^0\gamma$ followed by the decay of $\rho^0$ meson into a pair of pions \cite{Amelin:95}.
Using the angular distribution of pions it was possible to extract the events with the longitudinal polarization of $\rho^0$ meson and with the transverse polarization.
The following result was obtained for the elements of the $\rho^0$ meson polarization density matrix
$$\dfrac{\rho_{00}}{\rho_{11}}=3.9\pm 0.9\,(\mbox{stat.})\pm 1.0\,(\mbox{sys.})\,.$$
We use this value in the relation
\begin{equation}
b=\left|\dfrac{(1-a^2)F_{\rho\gamma}(m_\rho^2,0)+a^2G_{\rho\gamma}(m_\rho^2,0)}{a G_{\rho\gamma}(m_\rho^2,0)}\right|^2=\dfrac{2\rho_{11}}{\rho_{00}}=0.51\pm 0.18\,,
\end{equation}
where $a=m_\rho/M \approx 0.6$. Then it follows from Eq.~\eqref{rhogamma} that
\begin{equation}\label{g1g2}
\left|1-(1-a^2)\frac{g_1}{g_2}\right|^2=\frac{b}{a^2}=1.4\pm 0.5\,.
\end{equation}
We also predict the width $\Gamma_{\rho\gamma}$ of the $f_1 (1285)\to \rho^0\gamma$ decay, 
\begin{equation}
\Gamma_{\rho\gamma}=\dfrac{2\pi \alpha M m_\rho |g_2|^2}{9a^6\Gamma_{\rho \to ee}}(1+b)(1-a^2) \,,
\end{equation}
and compare this quantity with the experimental value $\Gamma_{\rho\gamma}=(1.2\pm 0.3)\,\mbox{MeV}$ \cite{pdg18}. As a result we obtain 
\begin{equation}\label{g22}
|g_2|=(2.9\pm 0.4)\cdot 10^{-4}\,.
\end{equation}
This value is in good agreement with \eqref{g21}. 

\begin{figure}[h]
\includegraphics[scale=1.0]{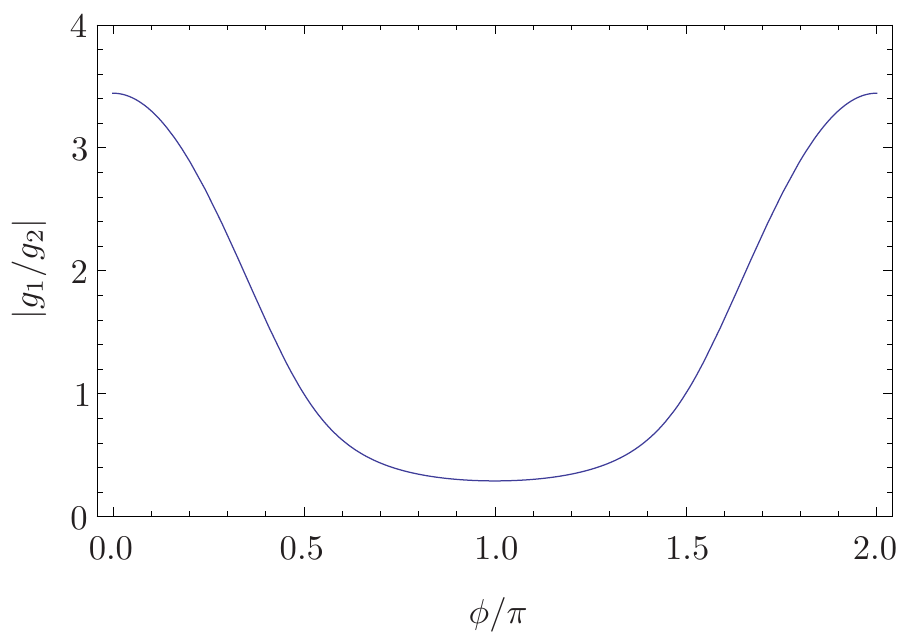}
\caption {The dependence of $|g_1/g_2|$ on $\phi$, Eq.~\eqref{gphi}}
\label{fig:g1g2}
\end{figure}

The ratio $|g_1/g_2|$ can be extracted from Eq.~\eqref{g1g2}. Since $g_1$ and $g_2$ are complex numbers, the value $|g_1/g_2|$ derived from \eqref{g1g2} depends on the relative phase 
$\phi$ of these numbers. Substituting $g_1/g_2= |g_1/g_2|e^{i\phi}$ in Eq.~\eqref{g1g2}, we obtain
\begin{equation}\label{gphi}
\left|\frac{g_1}{g_2}\right|=\dfrac{\cos\phi+\sqrt{b/a^2-\sin^2\phi}}{1-a^2}\,.
\end{equation}
 For instance,
\begin{equation}\label{g1g2ans}
|g_1/g_2|= 3.4\pm 0.3\quad \mbox{at}\ \; \phi=0\,,\quad 
|g_1/g_2|= 0.3\pm 0.3\quad \mbox{at}\ \; \phi=\pi\,. 
\end{equation}
The dependence of $|g_1/g_2|$ on $\phi$ is shown in Fig.~\ref{fig:g1g2}.

It is seen from Fig.~\ref{fig:g1g2} that the value $|g_1/g_2|$ is almost constant and small when $\phi$ varies in a wide region from $\pi/2$ to $3\pi/2$. That means that to fix $\phi$ together with $|g_1/g_2|$ it is necessary to use another data (see Section \ref{V}).

%---------------------------------------------------------

\section{Process $e^+e^-\to f_1 (1285)$}
Recently, the first observation of $f_1 (1285)$ meson production in $e^+e^-$ annihilation has been reported \cite{Achasov:19}. An estimate of the cross section of this process was made in 
\cite{Rudenko:17} for another parameterization of the form factors.
However, the predictions for the cross section of the process $e^+e^-\to e^+e^-f_1 (1285)$ following from that parameterization differ substantially from the experimental data \cite{Achard:02}. Therefore, it is interesting to compare the predictions for this process following from our parameterization of the form factors (see next Section).

\begin{figure}[h]
\includegraphics[scale=1.0]{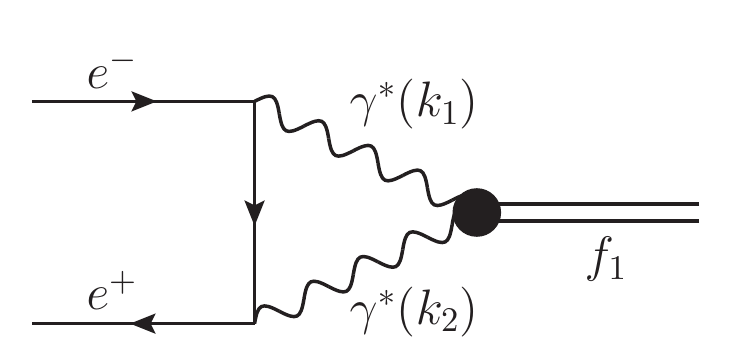}
\caption {The Feynman diagram of the process $e^+e^-\to f_1 (1285)$}
\label{fig:eef}
\end{figure}

Let us consider the process $e^+e^-\to f_1 (1285)$, the corresponding Feynman diagram is shown in Fig.~\ref{fig:eef}. 
Calculations are performed in the rest frame of $f_1 (1285)$ meson. We direct the $z$ axis along the electron momentum $\bm p$. 
Since electron and positron annihilate with the opposite helicities, the projection $\Lambda$ of the spin of $f_1 (1285)$ meson onto the $z$ axis can only be $\Lambda=\pm 1$.
Using Eq.~\eqref{ampl}, we derive the amplitude ${\cal T}_\Lambda$ for the process under discussion
\begin{equation}
\begin{gathered}
{\cal T}_\Lambda=(\bm A^*\cdot\bm e_\Lambda){\cal T}\,, \\
{\cal T}=\dfrac{i\sqrt{2}\alpha}{M^2} \int \dfrac{d\omega d^3q}{(2\pi)^4}
\Bigg\{\dfrac{1}{k_1^2}G(k_1^2,k_2^2)+\dfrac{1}{k_2^2}G(k_2^2,k_1^2)
+\dfrac{1}{D}\left[G(k_1^2,k_2^2)+G(k_2^2,k_1^2)\right] \\
+\dfrac{[\bm q\times\bm n]^2}{D}\left[\left(\dfrac{1}{k_1^2}-\dfrac{1}{k_2^2}\right)F(k_1^2,k_2^2)
+\dfrac{1}{k_1^2}G(k_1^2,k_2^2)+\dfrac{1}{k_2^2}G(k_2^2,k_1^2)\right]\Bigg\}\,.
\end{gathered}
\end{equation}

\begin{figure}[h]
\includegraphics[scale=1.0]{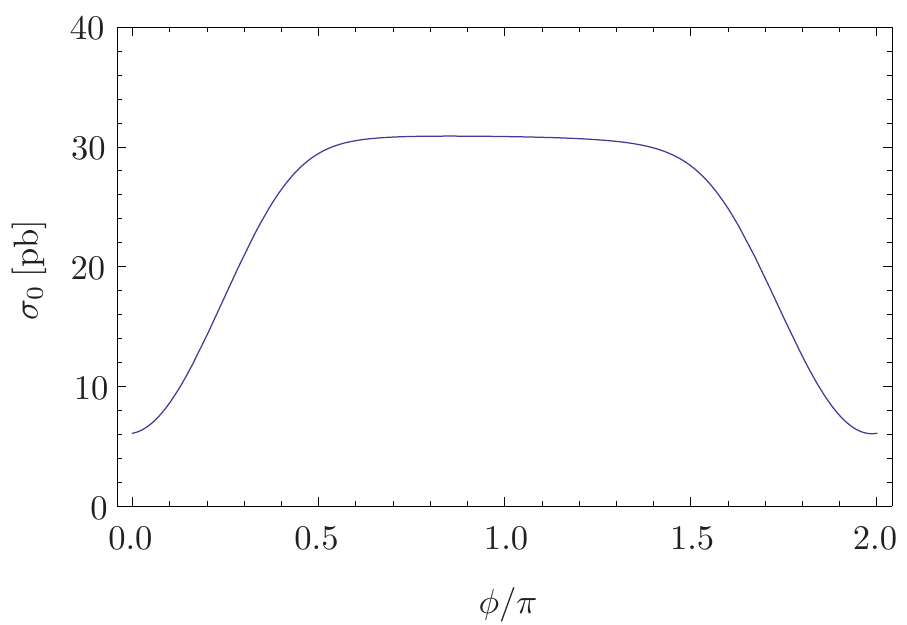}
\caption {The dependence of $\sigma_0$ on $\phi$, Eq.~\eqref{sig0}}
\label{fig:cross-section}
\end{figure}

The following notation is introduced in this formula:
\begin{equation}
\begin{gathered}
k_1^2=\omega^2-q^2+\omega M+\dfrac{1}{4}M^2\,,\quad k_2^2=\omega^2-q^2-\omega M+\dfrac{1}{4}M^2\,, \\
D=\omega^2-q^2-\dfrac{1}{4}M^2+2\bm q\cdot\bm p\,,\quad \bm n=\bm p/p\,,\quad \bm e_{\Lambda}=\dfrac{1}{\sqrt{2}}(-i\bm e_x+\Lambda\bm e_y)\,.
\end{gathered}
\end{equation}
Substituting our representation \eqref{gammagamma} of the form factors and performing integration, we obtain
\begin{equation}\label{final}
\begin{gathered}
{\cal T} = \alpha(c_1g_1+c_2g_2)\,, \\
c_1=0.41-4.76\, i\,,\quad c_2=-0.84+30.61\, i\,.
\end{gathered}
\end{equation}
It is seen that the imaginary parts of the coefficients $c_ {1,2}$ significantly exceed the corresponding real parts.
For unpolarized electron and positron beams, the cross section $\sigma_0$ at the peak is
\begin{equation}\label{sig0}
\sigma_0=\dfrac{1}{2M\Gamma_f}|{\cal T}|^2=3.56\,|c_1g_1+c_2g_2|^2\cdot\,10^{-7}\,\mbox{b},
\end{equation}
where $\Gamma_f= 22.7\,\mbox{MeV}$ is the $f_1 (1285)$ meson width. 
Using Eqs.~\eqref{g22}, \eqref{g1g2ans}, and \eqref{final} we obtain the prediction for the cross section $\sigma_0$ 
\begin{equation}
\sigma_0=(6\pm 2)\, \mbox{pb}\quad \mbox{at}\ \; \phi=0\,,\quad 
\sigma_0=(31\pm 16)\, \mbox{pb}\quad \mbox{at}\ \; \phi=\pi\,, 
\end{equation}
and compare it with the experimental result \cite{Achasov:19}
$$ \sigma_0=45^{+33}_{-24}\,\, \mbox{pb}\,.$$
Thus, we arrive at a conclusion that good agreement is achieved at $\phi=\pi$.
The dependence of $\sigma_0$ on $\phi$ is shown in Fig.~\ref{fig:cross-section}.

%---------------------------------------------------------

\section{Process $e^+e^-\to e^+e^-f_1 (1285)$}\label{V}

\begin{figure}[h]
\includegraphics[scale=1.0]{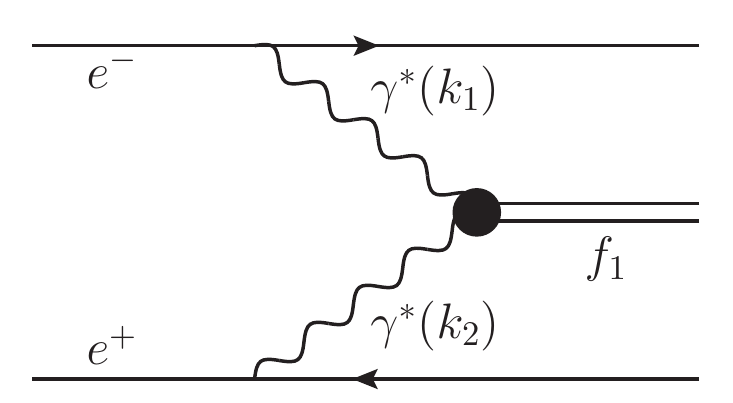}
\caption {The Feynman diagram of the process $e^+e^-\to e^+e^-f_1 (1285)$}
\label{fig:eefee}
\end{figure}

The Feynman diagram corresponding to the process $e^+e^-\to e^+e^-f_1 (1285)$ is shown 
in Fig.~\ref{fig:eefee}. 

It is convenient to describe this process in terms of the quantities 
$\sigma_{TT}$, $\sigma_{TS}$, $\sigma_{ST}$, $\sigma_{SS}$, $\tau_{TS}$, and $\tau_{TT}$, see~\cite{Schuler:98}.
These quantities are expressed via the spiral amplitudes ${\cal M}_{\lambda_1\lambda_2}$, 
Eq.~\eqref{ampl}, as follows
\begin{equation}\label{sigtau}
\begin{gathered} 
\sigma_{TT}=\dfrac{1}{4\sqrt{X}}[T(++,++)+T(+-,+-)]\,,\quad 
\sigma_{SS}=\dfrac{1}{2\sqrt{X}}T(00,00)\,, \\
\sigma_{TS}=\dfrac{1}{2\sqrt{X}}T(+0,+0)\,,\quad 
\sigma_{ST}=\dfrac{1}{2\sqrt{X}}T(0+,0+)\,, \\
\tau_{TS}=\dfrac{1}{4\sqrt{X}}[T(++,00)+T(-0,0+)]\,,\quad 
\tau_{TT}=\dfrac{1}{2\sqrt{X}}T(++,--)\,, \\
T(\lambda_1\lambda_2,\lambda_3\lambda_4) = \dfrac{M\Gamma_f}{(W-M^2)^2+M^2\Gamma_f^2}
\sum\limits_{\lambda_f}{\cal M}^*_{\lambda_1\lambda_2}{\cal M}_{\lambda_3\lambda_4}\,, \\
X=\nu^2-k_1^2k_2^2\,,\quad W=k_1^2+k_2^2+2\nu\,,\quad\nu=k_1k_2\,,
\end{gathered}
\end{equation}
where summation over $\lambda_f$ means summation over polarizations of $f_1 (1285)$ meson.
Then we obtain
\begin{equation} \label{sigsig}
\begin{gathered} 
\sigma_{TT}=\dfrac{N}{W}\Big|2XF(k_1^2,k_2^2)-k_2^2(k_1^2+\nu)G(k_1^2,k_2^2)+k_1^2(k_2^2+\nu)G(k_2^2,k_1^2)\Big|^2\,,\quad 
\sigma_{SS}=0\,, \\
\sigma_{TS}=2(-k_2^2)N\Big|k_1^2G(k_2^2,k_1^2)-\nu G(k_1^2,k_2^2)\Big|^2\,,\quad \\ \sigma_{ST}=2(-k_1^2)N\Big|k_2^2G(k_1^2,k_2^2)-\nu G(k_2^2,k_1^2)\Big|^2\,, \\
\tau_{TS}=\sqrt{k_1^2k_2^2}N\Big[k_1^2G(k_2^2,k_1^2)-\nu G(k_1^2,k_2^2)\Big]\Big[k_2^2 G(k_1^2,k_2^2)-\nu G(k_2^2,k_1^2)\Big]\,, \\ 
\tau_{TT}=-2\sigma_{TT}\,,\quad 
N=\dfrac{\Gamma_f}{4\sqrt{X}M^3[(W-M^2)^2+M^2\Gamma_f^2]} \,.
\end{gathered}
\end{equation}

\begin{figure}[h]
\includegraphics[scale=1.0]{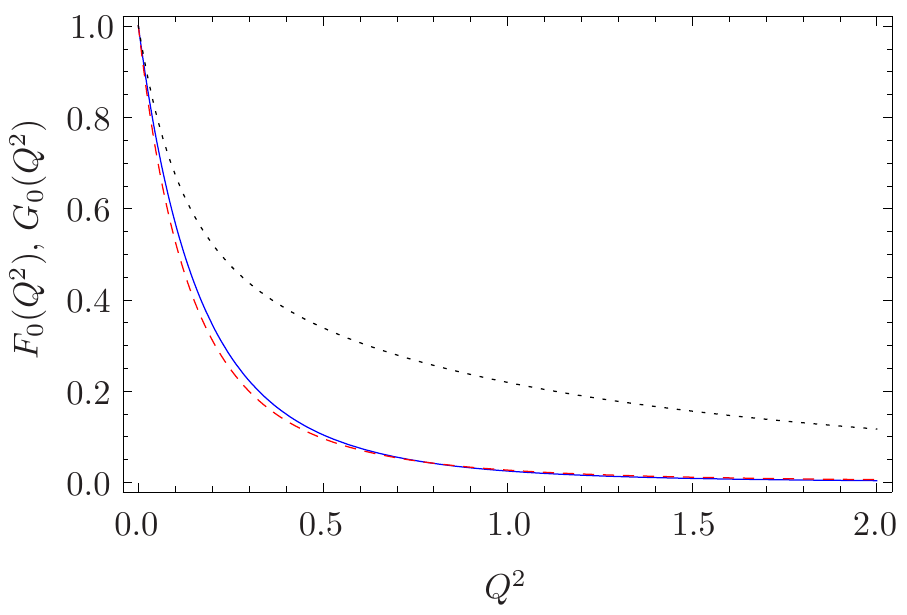}
\caption {Comparison of the functions $F_0(Q^2)$ (solid line), 
$G_0(Q^2)$ at $\phi=\pi$ (dashed line), and $G_0(Q^2)$ at $\phi=0$ (dotted line)}
\label{fig:compare}
\end{figure}

L3 Collaboration has studied a production of $f_1 (1285)$ meson for 
$k_1^2=0$, $k_2^2<0$ and $Q^2=-k_2^2\lesssim M^2$ \cite{Achard:02}.
The cross section $\sigma_{\gamma\gamma\to f_1 (1285)}$ in the peak was parameterized in \cite{Achard:02} as 
\begin{equation} \label{F0}
\sigma_{\gamma\gamma\to f_1 (1285)}=\dfrac{48\pi \widetilde\Gamma_{\gamma\gamma}}{M^2 \Gamma_f}(1+x)x\left(1+\frac{x}{2}\right)F_0(Q^2)\,,\quad x=Q^2/M^2\,,
\end{equation}
with the effective form factor 
$F_0(Q^2)=(1+Q^2/\Lambda_0^2)^{-4}$ 
and $\Lambda_0=1.04\pm0.06\pm0.05\,\mbox{GeV}$, \linebreak
$\widetilde\Gamma_{\gamma\gamma}$ is given in \eqref{gt}.
Using Eq.~\eqref{sigsig} we arrived at Eq.~\eqref{F0} with the replacement $F_0(Q^2)\rightarrow G_0(Q^2)$, where $G_0(Q^2)$ is
\begin{equation}
G_0(Q^2)=\dfrac{2+x|1-(1+x)g_1/g_2|^2}{(2+x)(1+x)^2(1+x/a^2)^2}\,,
\quad a=m_\rho/M\approx0.6\,.
\end{equation}
The comparison of $F_0(Q^2)$ and $G_0(Q^2)$ is performed in Fig.~\ref{fig:compare} where the function $F_0(Q^2)$ is shown as a solid line, $G_0(Q^2)$ at $\phi=\pi$ as a dashed line, and $G_0(Q^2)$ at $\phi=0$ as a dotted line. One can see a perfect coincidence of $F_0(Q^2)$ and $G_0(Q^2)$ at $\phi=\pi$. Thus, it follows from comparison of our predictions with the results of various experiments that the phase $\phi$ should be close to $\pi$. Therefore, the ratio $|g_1/g_2|$ is relatively small, $|g_1/g_2|= 0.3\pm 0.3$.

%---------------------------------------------------------

\section{Decays $f_1 (1285) \to \rho^0\pi^+\pi^-$ and $f_1 (1285) \to 2\pi^+ 2\pi^-$}
Withing our model, we can predict the widths $\Gamma(f_1 (1285)\to\rho^0\pi^+\pi^-)$ and $\Gamma(f_1 (1285)\to2\pi^+2\pi^-)$ of $f_1 (1285) \to \rho^0\pi^+\pi^-$ and $f_1 (1285) \to 2\pi^+ 2\pi^-$ decays and compare these predictions with the experimental data. It follows from the experimental results~\cite{Barberis:00} that the main contribution to the $f_1 (1285) \to 2\pi^+2\pi^-$ decay width is given by the $\rho^0\rho^0$ intermediate state. 
Neglecting the $\rho^0$ meson width in the $\rho^0$ meson propagator, we obtain from Eqs.~\eqref{rhorho} and \eqref{ampl}: 
\begin{equation}
\begin{gathered} 
\Gamma(f_1 (1285)\to\rho^0\pi^+\pi^-)=\frac{\alpha^2 |g_2|^2 M^2 \Gamma_\rho}{27 a^7 (1-4m_\pi^2/m_\rho^2)^{3/2}\,\Gamma^2_{\rho \to ee}}
\int\limits_{y_0}^{y_1} dy\,\dfrac{y}{Q}\left(1-\dfrac{4m_\pi^2}{M^2 y}\right)^{3/2} \\
\times 
\left[\left|4\dfrac{g_1}{g_2} Q^2-1+a^2+y\right|^2+\dfrac{y(y-1+3a^2)^2+a^2(3y-1+a^2)^2}{(y-a^2)^2}\right]\,, \\
Q=\dfrac{1}{2}\sqrt{[(1-a)^2-y][(1+a)^2-y]}\,,\quad y_0=\dfrac{4m_\pi^2}{M^2}\,,\quad y_1=(1-a)^2\,,
\end{gathered}
\end{equation}
where $m_\pi$ is the $\pi^\pm$ meson mass and $a=m_\rho/M \approx 0.6$. 
The value of $\Gamma(f_1 (1285)\to\rho^0\pi^+\pi^-)$ calculated by this formula is very sensitive to the value of the $\rho^0$ meson mass. For $m_\rho=775\,\mbox{MeV}$, we obtain the result which is almost two times smaller than the experimental value 
${\cal B}(f_1 (1285)\to\rho^0\pi^+\pi^-)=(11.2^{+0.7}_{-0.6})\,\%$ \cite{pdg18}. However, 
decreasing $m_\rho$ by only 20\% of the width $\Gamma_\rho$ we obtain ${\cal B}(f_1 (1285)\to\rho^0\pi^+\pi^-)=(8.7\pm 3.9)\,\%$ which is in good agreement with the experimental value \cite{pdg18}. Therefore, it is necessary to accurately account for the finite $\rho^0$ meson width. Note that a variation of $\phi$ from $\pi/2$ to $3\pi/2$, when $|g_1/g_2|$ is almost constant (see Section~\ref{P}), leads to a small variation of $\Gamma(f_1(1285)\to\rho^0\pi^+\pi^-)$, $\delta\Gamma/\Gamma\sim 1\%$.

Since it was shown in Ref.~\cite{Barberis:00} that the main contribution to the $f_1(1285) \to 2\pi^+2\pi^-$ decay width is given by the $\rho^0\rho^0$ intermediate state, so that ${\cal B}(f_1(1285)\to2\pi^+2\pi^-)={\cal B}(f_1(1285)\to\rho^0\pi^+\pi^-)$~\cite{pdg18}, our predictions also agree with the experimental data for $f_1(1285) \to 2\pi^+2\pi^-$ decay.

%---------------------------------------------------------

\section{Conclusion}

In this paper we propose parameterization of the $f_1 (1285)$ meson form factors and compare our predictions with the available experimental data. For $\phi$ close to $\pi$, where $\phi$ is the relative phase of the constants $g_1$ and $g_2$ in the form factors, we find good agreement between our theoretical predictions and the experimental data. 

\subsection*{Acknowledgments}
We are grateful to V.P.\,Druzhinin for useful discussions.


\begin{thebibliography}{99}

\bibitem{Landau:48}
L.D.\,Landau,
{\it On the angular momentum of a system of two photons},
Dokl. Akad. Nauk USSR Ser. Fiz. {\bf 60} (1948) 207;\\
C.N.\,Yang,
{\it Selection rules for the dematerialization of a particle into two photons},
Phys. Rev. {\bf 77} (1950) 242.

\bibitem{Gidal:87}
G.\,Gidal {\it et al}. (Mark II Collaboration),
{\it Observation of spin-1 $f_1 (1285)$ in the reaction $\gamma\gamma^* \to \eta^0 \pi^+ \pi^-$},
Phys. Rev. Lett. {\bf 59} (1987) 2012.

\bibitem{Aihara:88}
H.\,Aihara {\it et al}. (TPC/2$\gamma$ Collaboration),
{\it $f_1 (1285)$ formation in photon photon fusion reactions},
Phys. Lett. {\bf B\,209} (1988) 107;\\
H.\,Aihara {\it et al}. (TPC/2$\gamma$ Collaboration),
{\it Formation of spin one mesons by photon-photon fusion},
Phys. Rev. {\bf D\,38} (1988) 1.

\bibitem{Amelin:95}
D.V.\,Amelin {\it et al}. (VES Collaboration),
{\it Study of the decay $f_1 (1285) \to \rho^0 (770) \gamma $},
Z. Phys. {\bf C\,66} (1995) 71.

\bibitem{Achard:02}
P.\,Achard {\it et al}. (L3 Collaboration),
{\it $f_1 (1285)$ formation in two photon collisions at LEP},
Phys. Lett. {\bf B\,526} (2002) 269 [hep-ex/0110073].

\bibitem{Dickson:16}
R.\,Dickson {\it et al}. (CLAS Collaboration),
{\it Photoproduction of the $f_1 (1285)$ meson},
Phys. Rev. {\bf C\,93} (2016) 065202 [arXiv:1604.07425].

\bibitem{Achasov:19}
M.N.\,Achasov {\it et al}. (SND Collaboration),
{\it Search for direct production of the $f_1 (1285)$ resonance in $e^+e^-$ collisions},
Phys. Lett. {\bf B\,800} (2020) 135074 [arXiv:1906.03838].

\bibitem{Kopp:74}
G.\,K{\"o}pp, T.F.\,Walsh, and P.\,Zerwas,
{\it Hadron production in virtual photon-photon annihilation},
Nucl. Phys. {\bf B\,70} (1974) 461.

\bibitem{Kaplan:78}
J.\,Kaplan and J.H.\,K{\"u}hn,
{\it Direct production of $1^{++}$ states in $e^+e^-$ annihilation},
Phys. Lett. {\bf B\,78} (1978) 252.

\bibitem{Kuhn:79}
J.H.\,K{\"u}hn, J.\,Kaplan, and E.G.O.\,Safiani,
{\it Electromagnetic annihilation of $e^+e^-$ into quarkonium states with even charge conjugation},
Nucl. Phys. {\bf B\,157} (1979) 125.

\bibitem{Renard:84}
F.M.\,Renard,
{\it $1^{\pm +}$ resonances in $\gamma\gamma$ collisions},
Nuovo Cim. {\bf A\,80} (1984) 1.

\bibitem{Cahn:87}
R.N.\,Cahn,
{\it Production of spin 1 resonances in $\gamma\gamma$ collisions},
Phys. Rev. {\bf D\,35} (1987) 3342;\\
R.N.\,Cahn,
{\it Cross-sections for single tagged two photon production of resonances},
Phys. Rev. {\bf D\,37} (1988) 833.

\bibitem{Schuler:98}
G.A.\,Schuler, F.A.\,Berends, and R.\,van Gulik,
{\it Meson photon transition form-factors and resonance cross-sections in $e^+e^-$ collisions},
Nucl. Phys. {\bf B\,523} (1998) 423 [hep-ph/9710462].

\bibitem{Kochelev:09}
N.I.\,Kochelev, M.\,Battaglieri, and R.\,De\,Vita,
{\it Exclusive photoproduction of $f_1 (1285)$ meson off the proton in kinematics 
available at the Jefferson Laboratory experimental facilities},
Phys. Rev. {\bf C\,80} (2009) 025201 [arXiv:0903.5369].

\bibitem{Wang:17}
Y.Y.\,Wang, L.J.\,Liu, E.\,Wang, and D.M.\,Li,
{\it Study on the reaction of $\gamma p \to f_1 (1285) p$ in Regge-effective Lagrangian approach},
Phys. Rev. {\bf D\,95} (2017) 096015 [arXiv:1701.06007].

\bibitem{Wang:2017}
X.Y.\,Wang and J.\,He,
{\it Analysis of recent CLAS data on $f_1 (1285)$ photoproduction},
Phys. Rev. {\bf D\,95} (2017) 094005 [arXiv:1702.06848].

\bibitem{Rudenko:17}
A.S.\,Rudenko,
{\it $f_1 (1285) \to e^+e^-$ decay and direct $f_1$ production in $e^+e^-$ collisions},
Phys. Rev. {\bf D\,96} (2017) 076004 [arXiv:1707.00545].

\bibitem{pdg18}
M.\,Tanabashi {\it et al}. (Particle Data Group),
{\it Review of particle physics}, Phys. Rev. {\bf D\,98} (2018) 030001.

\bibitem{Barberis:00}
D.\,Barberis {\it et al}. (WA102 Collaboration),
{\it A spin analysis of the $4\pi$ channels produced in central pp interactions at 450 GeV/c},
Phys. Lett. {\bf B\,471} (2000) 440 [hep-ex/9912005].

\end{thebibliography}
\end{document}